# Thermoelectric Devices
## Cooling and Power Generation

Prof. Dr. Yehea Ismail    Ahmed Al-Askalany
Center of Nanoelectronics and Devices
The American University in Cairo
Cairo, Egypt
y.ismail@aucegypt.edu, a.alaskalany@aucegypt.edu

*Abstract* – **Thermoelectricity is the direct conversion of temperature gradient to electric voltage, and vice-versa. There are several potential applications of thermoelectricity, ranging from clean noiseless cooling, to waste-power harvesting in automotives. The performance of thermoelectric devices is still a challenge, and many approaches have been researched to overcome such obstacle. This paper seeks to give an overview of the thermoelectric phenomenon, the bulk semiconductor thermoelectric devices, and the new ways to increase performance of thermoelectric devices using new material structures and going to low-dimensional materials.**

## I. INTRODUCTION

Thermoelectricity is the direct conversion of temperature gradient to electrical potential difference. The phenomenon discovery dates back to 1800s, with many potential applications now and in the future, but also with a challenge to increase thermoelectric devices' performance. The phenomenon is described by three effects. The Seebeck effect -discovered by Thomas Seebeck- in 1821 is the conversion of temperature difference into electrical voltage. The Peltier effect -discovered by Jean Peltier in 1934- is the conversion of voltage into temperature gradient. And the Thomson effect –discovered by William Thomson in 1851- is the heat flow in a conductor with terminals held at different temperatures, due to current flow.

Thermoelectric devices can be used for heating/cooling purposes, or for power generation. The applications for thermoelectric devices can span different areas and industries. Waste-heat energy harvesting in automotives is one promising application for such technology [1]. Remote space missions use thermoelectric devices for power generation.

The main challenge for thermoelectric devices' design is the device performance, and the compromise between material parameters to increase it. The attempts to increase the performance of thermoelectric materials and devices are many, from material parameters optimization, to using going to new material structures, and low-dimensional thermoelement [1].

## II. THERMOELECTRIC EFFECTS

As mentioned in the introduction, Thermoelectricity is described by three thermoelectric effects.

The "Seebeck Effect" is the conversion of temperature gradient across the junctions of two dissimilar metals to electrical voltage in the range of millivolts per Kelvin difference. The effect is non-linear with temperature, and depends on absolute temperature, type and structure of materials. The Seebeck coefficient S is the amount of voltage difference ΔVgenerated for an applied temperature difference ΔT.

$$S = \Delta V / \Delta T \qquad (1)$$

The voltage difference V, then can be calculated using the following equation,

$$V = \int_{T_c}^{T_h} \left( S_B(T) - S_A(T) \right) dT \qquad (2)$$

The "Peltier Effect" is that a current flow causes a temperature gradient across the junctions of two dissimilar metals. The heat transfer is in the same direction of charge carriers. The thermal current density q is given by the following equation, where π and j are the Peltier coefficient, and electrical current density.

$$q = \pi\, j \qquad (3)$$

The "Thomson Effect" is the heat flow across a conductor, with terminals at different temperatures, due to current flow. The heat flow is given by:

$$\frac{dQ}{dx} = \mu I \frac{dT}{dx} \qquad (4)$$

The Thomson effect is the only measurable effect among the three thermoelectric effects, for a certain material, since the two other effects are related to pairs of materials.

Thomson/Kelvin relationship described in the following equations relates the three thermoelectric effects.

$$\pi = ST \qquad (5)$$

$$\mu = T \frac{dS}{dT} \qquad (6)$$

π, S, T, μ are the Peltier coefficient, the Seebeck coefficient, the absolute temperature, and the Thomson coefficient.

## III. THERMOELECTRIC FIGURE OF MERIT

The performance of a thermoelectric device is indicated by the figure of merit (FOM) ZT, it is a dimensionless quantity, which is given by:

$$ZT = \frac{GS^2 T}{K_l + K_e} \quad (7)$$

$G$, $K_l$, $K_e$ and T are electrical conductivity, lattice thermal conductivity, electronic thermal conductivity, and absolute average temperature.

For higher ZT, a large Seebeck coefficient and electrical conductivity, and small thermal conductivity are required. Thermal conductivity decreases the FOM, since it leads to undesired thermal exchange between hot and cold sides of a thermoelectric device.

The compromise to increase ZT is shown in fig. 1, Increasing the carrier concentration in order to increase electrical conductivity, also gives rise to thermal conductivity, and the Seebeck coefficient starts to decrease with higher thermal conductivity, due to lack of the ability of the material to maintain higher temperature gradient.

$Bi_2Te_3$ Is the most commonly used semiconductor material because of its relatively high FOM ZT=1 at 300°K.

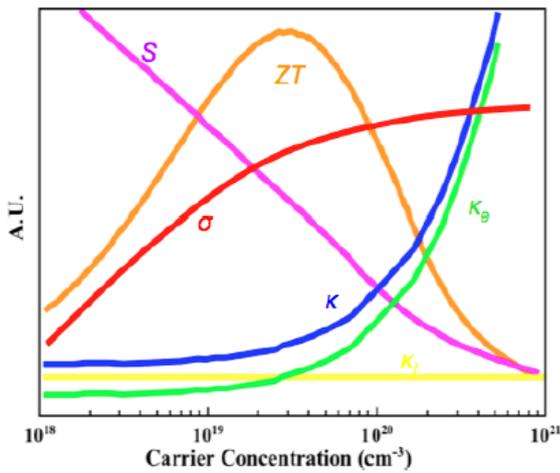

Figure 1: shows the FOM for some p-type semiconductor materials [1].

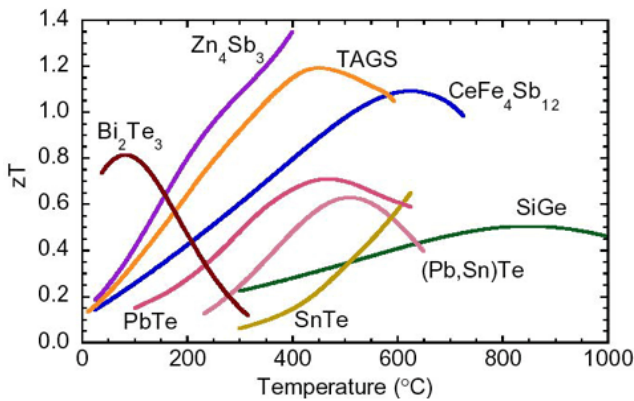

Figure 2: FOM for p-type materials [1].

## IV. BULK SEMICONDUCTOR TEC DEVICES

Thermoelectric cooling device (TEC), or Peltier cooler, depends in its operation on the Peltier effect described in the introduction. A current flow in a closed circuit of two dissimilar materials causes a temperature gradient across the junctions.

In bulk thermoelectric coolers, the semiconductor material is put between two conductor plates. The charge carriers are responsible for the process of heat transfer, due to the different energies required for their movement in the semiconductor material and the conductor material. The thermal current is in the same direction as charge carriers, as shown in fig. 3.

For an n-type thermoelement, electrons move freely in the conductor, and when they reach n-type material, they require more energy, which is absorbed from the surrounding environment. When moving from n-type back to the conductor, heat is released. For a p-type thermoelement, electrons entering semiconductor fill a hole, this drop in energy level is associated with heat release. And back to the conductor, electrons are back to a higher level of energy by absorbing heat.

For more heat flow more thermoelectric elements are required. Different approaches were implemented using single type arrangements. A configuration of multiple single-type thermoelements connected in parallel electrically and thermally, Fig. 4.a. This configuration requires low voltage with high current, which was impractical commercially. Another configuration was a set of single-type thermoelements connected parallel thermally and in series electrically, fig. 4.b. This would allow for larger heat transfer, but interconnects shorting the conductor surfaces limited the process of heat transfer.

The solution for the mentioned problem is using both n-type and p-type thermoelements to form a couple. This configuration is called a "Peltier Thermocouple", shown in fig 4.c.

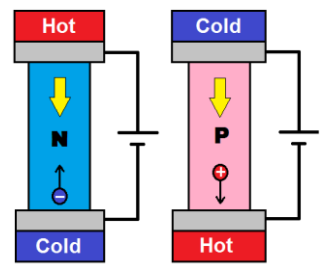

Figure 3: N-type and p-type thermoelements.

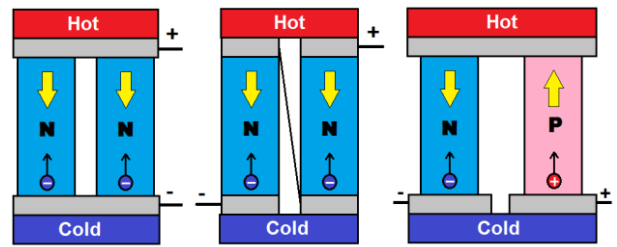

Figure 4: a) n-type parallel TEC. b) n-type series TEC. c) Peltier thermocouple.

## V. THE THERMOPILE

Using more thermocouples connected in series increases heat transfer. A set of alternating thermocouples are connected in series electrically and in parallel thermally. Such configuration is called the "Peltier Thermopile", fig. 5.

To find the different parameters of a thermopile, a set of equations are given [5], relating those parameters to the number of thermoelements in the thermopile and material parameters. The Seebeck coefficient of the total number n of thermocouples is given by,

$$S = n \int_{T_c}^{T_h} (S_p - S_n) dT \tag{8}$$

The device resistance and thermal conductance (of materials and interconnects) are approximated by,

$$R = n\left(\frac{\rho_n l}{A_n} + \frac{\rho_p l}{A_p} + R_l\right) \tag{9}$$

$$K = n\left(\frac{K_n l}{A_n} + \frac{K_p l}{A_p} + K_l\right) \tag{10}$$

The heat transferred is the sum of the Peltier, Fourier, and Joule heat terms,

$$Q = IST - K\Delta T - \frac{1}{2}I^2 R \tag{11}$$

When operating as a thermoelectric cooler. Maximum current, maximum heat pumping, maximum temperature difference are given by,

$$I_{max} = ST_c/R, \tag{12}$$

$$Q_{max} = S^2 T_c^2 / 2R, \tag{13}$$

$$\Delta T_{max} = ZT_c^2/2, \tag{14}$$

And the device's FOM is,

$$ZT = S^2 T / KR. \tag{15}$$

When operating as a thermoelectric generator. The efficiency is approximated with,

$$\eta = \frac{\Delta T}{T_h} \frac{\sqrt{1+ZT} - 1}{\sqrt{1+ZT} + T_c/T_h} \tag{16}$$

For a TEG, the efficiency increases with the increase in temperature difference, approaching Carnot efficiency limit.

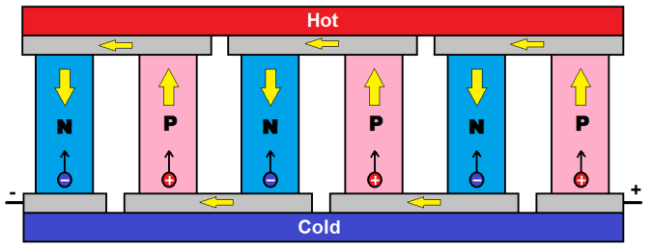

Figure 5: Peltier Thermopile.

Table 1: TEC1-12710 module parameters.

| Parameter | Value |
| --- | --- |
| Max Current | 10 Amps |
| Max Voltage | 15.4 Volts |
| Seebeck Coefficient | 0.0513 |
| Internal Resistance | 1.1909 Ohms |
| FOM (Z) | 0.0026 |
| Thermal Conductivity | 0.8757 |

## VI. THERMOELECTRIC COOLER MATLAB SIMULATION

In order to account for thermoelectric coolers in design, a model [3] of the cooler should be made, relating its electrical and thermal characteristics to manufacturer's data sheet for the device. A model that defines the required current to achieve certain temperature difference is given as follows,

$$I = \frac{S(T_H - \Delta T) - S\sqrt{(T_H - \Delta T)^2 - 2\Delta T/Z}}{R} \tag{17}$$

And the Voltage required,

$$V = V_{max} - S\sqrt{(T_H - \Delta T)^2 - 2\Delta T/Z} \tag{18}$$

Given the manufacturer's data for the thermoelectric cooler "TEC1-12710" in the table 1[3], A Matlab code was written to demonstrate the current and voltages required for a range of temperature differences.

The hot side temperature was kept constant at 300°K, and varying the cold side temperature from 230°K to 300°, and plotting the required current and voltage versus cold side temperature.

The results shown in fig. 6 shows that the amount of current/voltage required is inversely proportional to the temperature difference desired, between the hot and cold sides of the device.

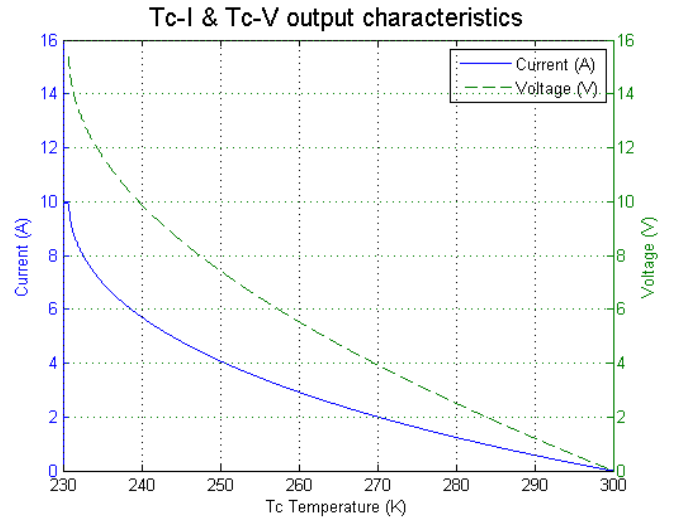

Figure 6: Tc-I and Tc-V of TEC model.

## VII. NEW MATERIALS FOR THERMOELECTRIC DEVICES

The use of quantum well structures and thin films for thermoelectric devices is the current research efforts in order to increase the performance of thermoelectric materials and devices. For example, alternating layers of Sb2Te3 and BiTe3 are used to make thin films used in thermoelectric coolers. Such new material dimensions and arrangements allowed for power densities up to $100W/cm^2$.

The use of thermoelectrical wires in the nanometer scale grown inside of a nanoporous aluminum matrix is used in order to improve electrical and thermal parameters beyond bulk materials.

Structures like alternating thin layers of quantum wells are used, which enhances the electrical and thermal parameters of the material. A device that consists of alternating layers of Si (barrier) and SiGe (quantum well) is shown in fig. 7. The structure was simulated using SILVACO TCAD, showing improved characteristics over bulk semiconductor thermoelectric devices [1].

The use of quantum well in semiconductor thin-layer materials affects materials parameters, the parameters of such structures is different from bulk materials, showing improvement in thermoelectric performance.

Figure 8 shows the effect of doping level and quantum well thickness on the Seebeck coefficient of the material. And fig. 9 shows the effect of doping level and well thickness on the numerator of ZT (FOM). Fig. 10 shows the change in thermal conductivity with size in ultra-thin crystal Si. Fig. 11 shows the effect of doping level and barrier thickness on the conductivity.

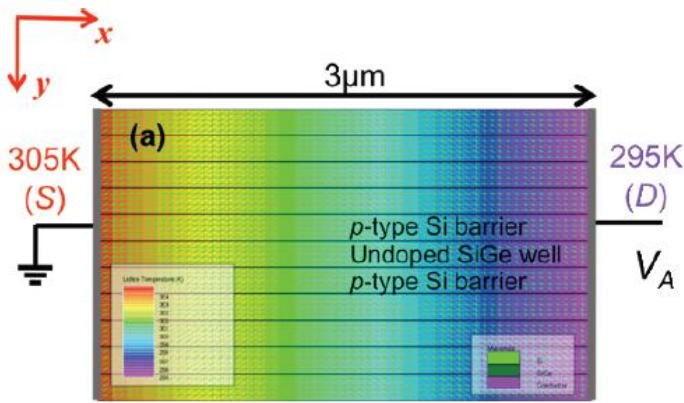

Figure 7: Si/SiGe Quantum-well thin film thermoelectric device [1]

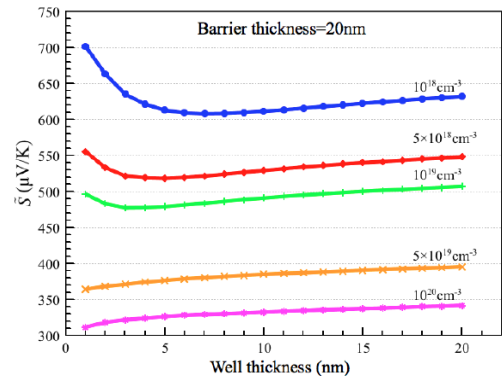

Figure 8: Doping and WT vs. S [1]

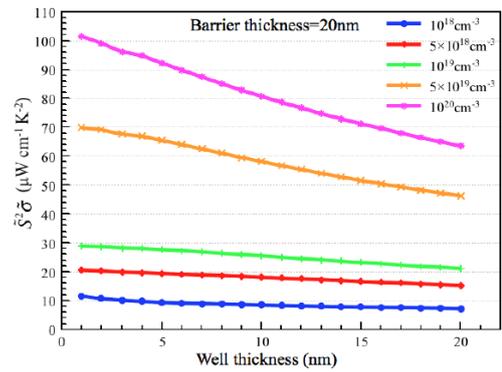

Figure 9: Doping and WT vs. ZT [1]

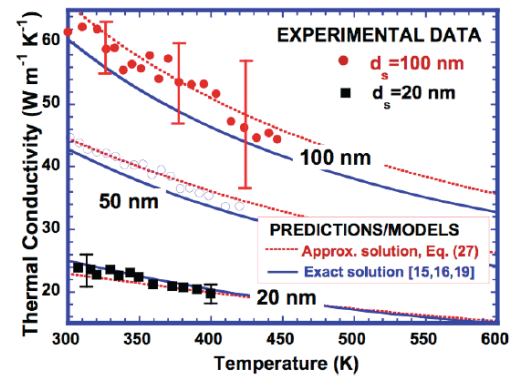

Figure 10: Size vs. Thermal conductivity [1]

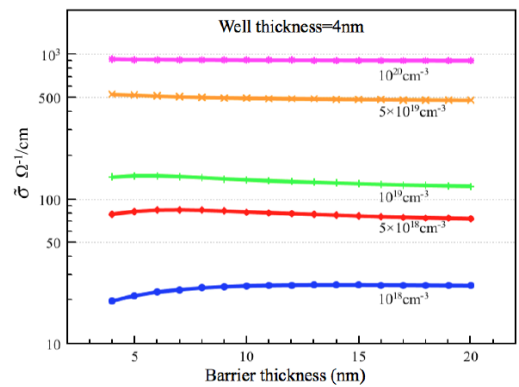

Figure11: Barrier thickness vs. electrical conductivity [1]

## VIII. Conclusion.

Thermoelectric devices have great potential for many applications, in different industries and fields. The thermoelectric coolers offer an environment friendly and very silent solution, it could be used in cars, homes, and food industry. Thermoelectric generators can be utilized in power harvesting in automotives, were excess thermal waste-energy is found.

The physics of thermoelectric devices is well known to us, and the tradeoff and compromise in increasing the performance of thermoelectric materials is not an easy task, the thermal and electrical parameters of a thermoelectric material are strongly entangled, and the seek for optimization of on parameters, directly affects another parameter in an undesired way.

Bulk semiconductors were firstly and still used for the manufacturing of thermoelectric devices; many assemblies were thought of and tried. The thermopile which consists of alternating thermocouples of n-type and p-type thermoelents is the best way to have larger assemblies of bulk semiconductor thermoelectric devices. The FOM of merit of thermoelectric cooler and the efficiency of thermoelectric devices not only depend on material parameters and temperature difference, but also on the absolute average temperature the device operates in, making different materials suitable for different operating temperatures.

The introduction of nanowires thermoelectric wires and structures with quantum wells has shown a promising improvement in the performance of such new devices.

Future research investigating the integration of low-dimensional material thermoelectric devices with solar energy materials would lead to a solution for the thermal waste-energy generated in solar power harvesting devices. Even if it could be possible to design a device that operates for both purposes, solar and thermal power harvesting, this would be great to compensate for the low power efficiency of both types of devices.

The work in simulation of quantum well thermoelectric devices using TCAD will be my starting point to get a deeper understanding of the tradeoffs in material parameters. And it will offer me an opportunity for future research in the subject, with the intention to find new ideas in materials and device design.